\def\gtorder{\mathrel{\raise.3ex\hbox{$>$}\mkern-14mu
             \lower0.6ex\hbox{$\sim$}}}
\def\ltorder{\mathrel{\raise.3ex\hbox{$<$}\mkern-14mu
             \lower0.6ex\hbox{$\sim$}}}

\def\euve{{\it EUVE }}
\def\rosat{{\it Rosat }}
\def\etal{et al. }

\documentstyle{mn}
\input{epsf.sty}
 \begin{document}
\title
{Evidence for a New Class of Extreme UV Sources}

\author[Dan Maoz, Eran O. Ofek, and Amotz Shemi]
      {Dan Maoz\thanks{E-mail address: dani@wise.tau.ac.il},
      Eran O. Ofek, and
       Amotz Shemi\\
       School of Physics and Astronomy and Wise observatory,
 The Raymond and Beverly Sackler Faculty of Exact Sciences,\\
       Tel-Aviv University,
       Tel-Aviv 69978, Israel}
\maketitle
\begin{abstract}
Most of the sources detected in the extreme ultraviolet (EUV; 100 \AA\ to 600 \AA)
by the {\it Rosat/WFC} and {\it EUVE} all-sky surveys have been identified
with active late-type stars and hot white dwarfs that are near enough
to escape absorption by interstellar gas. However, about 15\% of EUV sources are 
as of yet unidentified with any optical counterparts. We examine
whether the unidentified EUV sources may consist of the same population
of late-type stars and white dwarfs. We present 
$B$ and $R$ photometry of stars in the fields of seven of the unidentified
EUV sources. 
We detect in the optical the entire main-sequence and white-dwarf
 population out to the greatest distances where they could still avoid absorption.
We use colour-magnitude diagrams to demonstrate that, in most of the
fields, none of the observed stars have the colours and magnitudes of
late-type dwarfs at distances less than 100 pc. Similarly,
none are white
dwarfs within 500 pc that are hot enough to be EUV-emitters.
 The unidentified EUV sources we study
are not detected in X-rays, while
cataclysmic variables, X-ray binaries, and active galactic nuclei generally are.
We conclude
that some of the EUV sources may be a new class of nearby objects, that are
either very faint at optical bands or which mimic the colours
and magnitudes of distant late-type stars or cool white dwarfs.
 One candidate for optically faint objects
is isolated old neutron stars, slowly accreting interstellar
matter. Such neutron stars are expected to be abundant in the Galaxy,
and have not been unambiguously detected.
\end{abstract}

\section{Introduction}
The {\it ROSAT Wide Field Camera
(WFC)} and the 
{\it Extreme Ultraviolet Explorer (EUVE)} missions carried out
all-sky surveys at extreme-ultraviolet (EUV) 
wavelengths in 1990-91 and 1992-93, respectively (Pounds \etal 1993;
Pye et al. 1995; Malina 
\etal 1994; Bowyer \etal 1994, 1996).
 The \euve all-sky
survey was carried out in four broad bands, centred at approximately
100 \AA, 200 \AA, 400 \AA, and 600 \AA. Simultaneously, a ``Deep Survey''
with one order of magnitude greater sensitivity was carried out
in a $2^\circ \times 180^\circ$ strip of sky along the ecliptic
in the 100 \AA\ and 200 \AA\ bands.
 The \rosat WFC all-sky survey was
conducted in two EUV bands, centred at about 100 \AA\ and 150 \AA.
 The 100 \AA\ bands in the two all-sky survey experiments
 had comparable sensitivities.

About 500 EUV sources
were detected by each of the surveys, and most of the brightest 300, or so,
sources were detected
mutually by \euve and \rosat (Barber et al. 1995). In ongoing efforts
by several groups, about 90\% of the \rosat EUV sources
(Mason et al. 1995) and 65\% to 90\% (depending on the sample definition)
of the \euve sources (Bowyer et al. 1996; 
 Craig et al. 1996b) have been optically identified.

 Among the identified sources,
about 55\% are late-type main-sequence dwarfs with chromospheric activity, and
35\% are hot white dwarfs. The remaining 10\% are cataclysmic variables (CVs), bright
early-type stars, and about 20 bright (13--14 mag) 
active galactic nuclei (AGNs; Barber et al. 1995; Mason et al. 1995;
Marshall, Fruscione, \& Carone 1995).

Warwick et al. (1993) showed that the \rosat EUV sources which have been
identified as white-dwarfs are distributed
within the $\sim 100$ pc-radius ``Local Bubble'' of low-density gas
   in the interstellar medium (ISM). More distant
white dwarfs are undetected in the EUV due to absorption by the
higher-density ISM gas. Among the
identified active late-type stars, the number/EUV-flux relations fall
off as expected for a homogeneously distributed population. In other 
words, the EUV detection of late-type stars is limited by their faint
EUV flux, before they reach the distance
 of the walls of the Local Bubble. The most distant EUV sources identified 
as late-type stars are at 90 pc, for F-type stars, and the distance decreases as one
goes to later, less luminous, types, to 40 pc
for M-dwarfs. Most EUV-detected M-stars are at about 10 pc (see Warwick
et al.1993). It cannot be presently excluded, however, that some of this
dependence of maximum detected distance on spectral type is a selection
effect of the optical identification programs, which generally concentrate
on the brightest objects in a field. For example, 
EUV sources associated with distant M-stars would more likely remain
unidentified than F-stars at the same distance, which are brighter.
 In any case,
even if some late-type stars were bright enough to be detectable in the EUV
at larger distances, they would be obscured in the EUV beyond $\sim$ 100 -- 200
pc, as evidenced by the limiting distances to the identified white dwarfs. It has been
speculated that many of the unidentified sources are faint white dwarfs
(Warwick et al. 1993; Barber et al. 1995) or late-type stars (Pye et al. 1995).

Most of the ongoing EUV-source identification programs proceed by obtaining
optical spectra of the brightest objects within the positional error-circles 
of the EUV sources.
These programs are ``positive'' in the sense that they
aim to complete the identification of as many EUV-sources as possible.
EUV sources with optical counterparts fainter than
the spectroscopy brightness limit will remain unidentified. 

In this paper, we describe the results of an identification program
with a ``negative'' orientation. We have subjectively selected among 
some of the unidentified EUV sources those that {\it a priori} appear
most difficult to identify. We have chosen those sources that lie in
relatively sparse and faint star fields. Our goal is to see whether
any of the unidentified EUV sources {\it cannot} be identified with
the common types of optical counterpart. Thus, while the sub-sample
of EUV fields we study does not represent statistically any part of
the EUV-source population, our program has the potential of identifying
new classes of objects with some of the EUV sources (or of showing that
such new classes are not yet required). 

An exciting possibility for a new class of objects producing some
of the unidentified EUV detections is 
nearby isolated old neutron stars (i.e. neutron stars not in binaries or in
supernova remnants, and too old to emit radiation as normal pulsars) accreting
material from the ISM (Ostriker, Rees \& Silk 1970; Treves \& Colpi 1991;
 Blaes \& Madau 1993; Shemi 1995a). 
The Galaxy is estimated to have $\sim 10^9$ isolated old neutron stars,   
a population 3-4 orders of magnitude larger than the pulsar population, but
which has received little attention. The closest isolated neutron star is
likely to be less than $\sim 10$ pc from Earth, and several hundreds are
expected within 100 pc.
 Discussions of the 
possible spectral
properties of isolated neutron stars appear in  
 Treves \&  Colpi (1991), Blaes \& Madau
(1993), Madau \& Blaes (1994), Zampieri \etal (1995), Nelson \etal (1995),
and Shemi (1995a, 1995b). Although the actual spectrum of such objects
is not clearly known yet, it is possible that they could be bright enough
in the EUV range to be detected by \rosat and \euve, but exceedingly
faint at optical energies (Shemi 1995a). Indeed, there have recently been
reports of two X-ray sources that may be candidates for this class
 (Stocke et al. 1995; Walter, Wolk,\& Neuhauser 1996).

In \S 2, below, we describe our optical observations of a selection of
the EUV source fields, and their reduction. In \S 3 we  present
 colour-magnitude diagrams for the stars in each field and discuss
whether any candidates for optical counterparts exist. We summarize
our results in \S 4.

\section{Observations and Reduction}

For our observations, we selected seven unidentified EUV
sources from the compilations of Pounds et al. (1993)
 and Bowyer et al. (1994, 1996).
 Apart from the criterion that the fields could be observed from the
Northern hemisphere, we generally chose fields that had
strong EUV detections, or were detected in more than one
band, or with both \rosat and \euve. As discussed above,
we generally chose sources in sparse fields devoid of
 bright stars, based on examination of Palomar Sky Survey
plates. However, some of the fields are fairly crowded.
As noted in more detail for the individual objects in \S 3, some
of the fields we study are not included in the latest
compilations of EUV sources detected by \rosat (Pye et al. 1995)
and \euve (Bowyer et al. 1996). At least one of these (EUVE $2114+503$,
and probably also EUVE $0807+210$)
has been erroneously omitted from
the catalogs. The others may have been false 
EUV detections, but they could also be real, and
simply below the significance threshold that was set for inclusion
in the latest catalogs.

We have searched the recent \rosat X-ray All-Sky Survey Bright
Source Catalog (Voges et al. 1996), which was conducted at 0.1--2.4 keV
 simultaneously
with the \rosat WFC survey, for an X-ray signal coincident to $5'$
with the positions of the unidentified
EUV sources. Two of the fields were detected in X-rays, EUVE $0807+210$
and EUVE $2053-175$. As it happens, these are the two fields in which
our analysis, below, points to late-type star candidates, and the
candidate positions correspond well to the X-ray source positions.
The other five EUV sources were not detected in the X-ray All-Sky Survey.
A search of the HEASARC archive of sources from other X-ray missions
 also does not turn up any documented
X-ray sources corresponding to our EUV sources 
 Table 1 lists the fields we have studied, their EUV
source parameters, and details of the optical
observations.

Observations were carried out with the Wise Observatory 1m
telescope and a Tektronix $1024\times 1024$-pixel
back-illuminated CCD.
 The pixel scale was $0.7''$ per pixel
in the direct imaging mode at the Cassegrain focus of the telescope,
or $2.08''$ per pixel when imaging with the FOSC instrument (which
for the present purposes is simply a focal re-imager).
Images of each field were obtained through standard Johnson-Cousins $B$
and $R$ filters. Photometric standard-star fields from Landolt (1992)
were also observed throughout each night. Several of the fields
were observed on non-photometric nights and were later calibrated
by means of brief exposures of the same fields on photometric nights.
The photometric solutions yield errors in the terms of the 
photometric calibration of order 0.01 mag, and an intrinsic scatter
of 0.01 to 0.03 mag. 

The CCD images were reduced in a standard way using the IRAF\footnote{IRAF (Image Reduction and Analysis
Facility) is distributed by the National Optical Astronomy Observatories,
which are operated by Aura, Inc., under cooperative agreement with the
National Science Foundation.} package.
The positions of the stars on the optical image were determined
with the astrometry task COORDS. The DAOPHOT task (Stetson 1987) was then run
on each image to automatically detect and measure the magnitudes
of all stars within a $3.5'$ radius of the EUV detection.
 The various sources of positional error in the \euve
and \rosat measurements are discussed by Craig et al. (1996b), Bowyer
et al. (1996), Pounds
et al. (1993), and Pye et al. (1995). The \rosat 90\%-confidence error circles
are listed for each source in the \rosat catalogs and are typically $1'$.
 The \euve 90\%-confidence error circles
are up to $1.4'$ for objects from the all-sky survey, and
up to $2.1'$ for Deep Survey sources. By considering
as candidates all stars within $3.5'$ of the EUV detection,
we are being highly conservative.
 Uncertainties in the final $B$ and $R$ magnitudes were
found by combining in quadrature the uncertainties in the photometric
solution, the scatter of the standard-star measurements around
the photometric solution, and the DAOPHOT uncertainties in the
PSF fitting of individual stars.
 
For the typical 15 min exposures, the automatic star detection
 is complete to stars of at least $B=21$ mag,
 for which total errors are 0.1 to 0.25 mag. The $R$ images
go to $R=22$ mag at this level of accuracy.
Some of the exposures were shorter or longer (see Table 1), and the
detection limits correspondingly shallower or deeper.
In all the
fields we have studied, all stars that were detected in $B$ were
also detected in $R$. On the other hand, many stars are detected
in $R$ but not in $B$. In the subsequent
analysis we will consider only stars detected in both the $B$ and the $R$
exposures. As we will show in \S 3, the faint red stars that do not fulfill
this condition are unlikely candidates as optical counterparts
to the EUV sources. 

Tables 2-8 (provided in microfiche form)
 list, for each field we have analysed, the $B$ 
and $R$ magnitudes and positions of all the stars in order
of increasing distance from the EUV source position.
The positions are typically accurate to better than $0.3''$
and can be used, e.g., for planning multi-slit spectroscopy
of these fields. The tables are also available electronically
on request from the authors.

\section{Analysis}
\subsection{Method}
In this section, we use colour-magnitude diagrams to 
search for optical counterparts to the EUV sources
 among the stars we have observed. As explained in \S 1,
the majority ($\sim 90\%$)
 of {\it identified} EUV sources are 
either nearby late-type main-sequence stars
(including RS CVn-type stars) or
 nearby white dwarfs. The remaining
10\% of identified EUV sources are composed of 
CVs (5\%), and a handful of 
early-type stars, X-ray binaries,
and AGNs. For each of these populations, we discuss below 
whether they could be responsible for the
unidentified EUV sources we study, and describe the region of the $R$ vs.
$B-R$ diagram they occupy. We will
then argue that, in fields where none of the stars are in 
the regions of the colour-magnitude diagram populated by 
the main known classes of EUV optical counterparts, another class of
object must be responsible for the EUV flux.

\subsubsection{Active Late-Type Stars}
Active late-type dwarfs nearer than $\sim 100$ pc constitute
the majority of EUV sources.
To determine the
region of the colour-magnitude diagram occupied by them,
the relation between absolute magnitude $M_R$, and $B-R$
colour for main-sequence dwarfs is required. For dwarfs of type 
A0 to M1, the $B-R$ colour-magnitude relation is relatively well
known. From Allen (1973) we derive the approximate empirical
relation:
$$
M_R=2.8(B-R)+0.7, \ \ \ \ \ \ \ \ \ 0\le(B-R)\le 2.8.
$$
For types M1 to M4, we estimate from Kirkpatrick et al. (1994):
$$
M_R=6.9(B-R)-10.78, \ \ \ \ \ \ \ \ \ 2.8\le(B-R)\le 3.3.
$$
For the lowest-mass stars, the relation between magnitude
and colour is subject to some uncertainty. In particular, there
is little information in the $B$ band, since studies of the 
faintest and reddest stars are naturally done in the red and infrared
parts of the spectrum. However, from Berriman \& Reid (1987) we see that
for the reddest stars, $(B-R)\approx (V-I)$, up to $(V-I)=4.5$.
(The reddest known stars have $V-I=4.7$ [Monet et al. 1994] or
$V-I\approx 5$ [Kirkpatrick et al. 1994].) Combining this rough equivalence
with the relations between $V,R$, and $I$ in Kirkpatrick et al. (1994),
we estimate for types M4 to M8:
$$
M_R \approx 3(B-R)+2.1, \ \ \ \ \ \ \ \ \ 3.3\le(B-R)\le 4.5.
$$

In addition to the present uncertainty in these relations, real stars have
an intrinsic spread about the relations of order 1-2 magnitudes.

Figures 1-7 show, for every field, an observed colour-magnitude
diagram of $R$ vs. $B-R$ for all the sources within $3.5'$ of the 
EUV source.  Also shown (parallel solid and dotted
curves) are the colour magnitude relations for a zero-age main sequence
(ZAMS) at various distances, as approximated above.
At the bottom of the plots we show (marked dashed lines) the $B$ and $R$ detection limits
for the exposures of each field. As seen in the figures, the detection limits
generally allow us to see all stars to the end of the main sequence out to a few tens pc.
At a few 100 pc, we can detect only dwarfs earlier than M. However, Warwick et al. (1993)
show that all EUV-sources identified to date with M stars are closer than 40 pc.
Since stars have an intrinsic spread of 1-2 magnitudes about the color-magnitude
relation, and late-type stars need to be closer than 100 pc to be detected in the EUV,
stars that are more than several magnitudes below the 100-pc  main sequence cannot
be late-type star EUV sources. Similarly, stars that are undetected in $B$
(and hence not plotted in the figures) lie below the diagonal line indicating
the $B$ detection limit, and so are not plausible late-type star EUV sources.

Note that any late-type stars detected in the EUV would also be expected to be
detected in even a relatively short expoure with the \rosat X-ray telescope.
The fact that our EUV sources are not detected in the \rosat All-Sky-Survey
in X-rays (save two EUV sources, which we indeed associate with late-type stars),
is an independent argument that our sources are not late-type stars.
(This argument does not necessarily hold for white dwarfs, to to their rather
``soft'' spectra).
 
\subsubsection{White Dwarfs}
The second population that is a main source of EUV detections is hot white dwarfs.
The short curves on the left side of the plots show the $R$ vs. $B-R$ relation for
white dwarfs at various distances, based on the relation given by Dawson (1986),
for white dwarfs with $B-R\le 0$. Figure 8 shows the distribution of $B-V$
colour of all white-dwarf EUV sources that have been detected by \rosat or \euve
with available absolute magnitudes. The absolute magnitudes are taken
from the compilation by McCook and Sion (1987), and the relation
between $M_V$ and $B-V$ is according to Dawson (1986). All white dwarfs that have been detected by
either satellite have $B-V\le 0.13$, which corresponds to $B-R\le 0$ (Dawson 1986).
White dwarfs with $B-R> 0$ are apparently too cool ($T\ltorder 1\times10^4$K)
 to be detected in the EUV.
Stars that are much redder than $B-R=0$ cannot be white-dwarf EUV sources.
As seen in Figs. 1--7, the detection limits of our observations allow us to
see the hot white dwarf population out to $\sim 500$ pc.

\subsubsection{Cataclysmic Variables}
CVs constitute about 5\% of identified EUV sources.
Examination of the properties of CVs that have been identified as EUV sources
shows that they have a variety of colours and absolute magnitudes. We therefore
cannot exclude that a CV is the EUV source in a given field
based solely on the colours and magnitudes of the stars in the field.

However, we note that most of the EUV sources that have been identified
with CVs are also X-ray sources. This is seen in Table 9, which
 lists all CV EUV sources detected by the \euve and \rosat surveys
and, in the right-hand column, the detection in the \rosat X-ray all-sky
survey (Voges et al. 1996).
Of the 23 CVs
detected by either of the EUV surveys 21 are X-ray sources. On the other hand,
among the EUV sources we have studied there are 3 \rosat X-ray sources:
EUVE $0807+210$ and EUVE $2053-175$ which we identify below as possible
 late-type-star EUV sources;
and RE 0922+71, where a background quasar unrelated to the EUV source
is reponsible for the X-ray emission (see Maoz et al. 1996, and below).
The remaining, unidentified, EUV sources are not X-ray sources.
This argues that it is unlikely that there are CVs behind all of the
unidentified EUV sources.
Nevertheless, since there are two examples of X-ray quiet CVs (which were
probably active during the \euve survey but quiescent during the \rosat
surveys), and these have
unremarkable colours and magnitudes, we cannot rule out that our EUV sources
are CVs.

\subsubsection{Other Sources}
About 15\% of late-type-star EUV sources that have been identified are
chromospherically or coronally active binary systems
 of the RS CVn and BY Dra-type (Pounds et al. 1993).
 All the RS-CVn cases, however, are
 brighter than $V=10$ mag, including some that are very faint in the EUV.
Identified BY-Dra binaries are all brighter than $V=11$ mag, but have red colours,
so they also fall above the 100 pc ZAMS on the colour magnitude diagram. 
Stars that cannot be main-sequence
 late-type-star EUV sources can therefore neither be binaries of these types
and be the origin of the EUV flux.
Binary systems consisting of a white dwarf and a late-type star can also
be EUV sources (Barstow et al. 1994). The EUV emission is dominated
by the white dwarf, while the optical light is primarily from the
late-type star. We therefore automatically test for sources of this
type when we test for late-type-star EUV sources.
Several X-ray binaries have been detected in the EUV. Since none of the 
unidentified EUV
sources we study in this paper are \rosat All-Sky-Survey 
X-ray sources, objects of this class
do not constitute EUV-source candidates.
About 20 bright (13--14 mag) 
AGNs have been detected in the EUV (Barber et al. 1995;
Marshall, Fruscione, \& Carone 1995). We have verified  that known AGNs are
not responsible for the EUV sources studied here. Note also that
AGNs have been detected in the EUV only at 100 \AA, and these are then
 always also X-ray sources.
Extragalactic objects are not detected at longer EUV wavelengths due to the
strong absorption by the ISM of the Galaxy.\footnote{Mild exceptions
to this statement are the marginal detections of 1H 1430+423 and
PKS 2155$-$304, both bright and variable BL Lac objects,
 in the \rosat 150 \AA\ band (Pye et al. 1995)}

\subsubsection{Absorption and Reddening}
The gas associated with any
 reddening by dust that affects the position of a star on the colour-magnitude
diagram by more than 0.05 mag in $B-R$ would effectively absorb all the EUV 
radiation from a source. Therefore, if there is a field with no candidate
optical counterparts we need not worry whether the colours and magnitudes of
some of the stars have been affected by dust. Conversely, fields that do
have very red but bright stars could be simply fields with high extinction,
rather than fields with nearby late-type stars that are the EUV sources.

\subsection{Individual Fields}
We examine below the colour-magnitude diagram for each of the fields individually.
For a star in one of our fields to be a member the same population
associated with most EUV sources,
it must be either above the ZAMS--100 pc
curve (i.e. a late type dwarf closer than 100 pc), or bluer than $B-R=0$ (i.e.
a hot white dwarf).

\subsubsection{EUVE 0715$+$141}

This EUV source from the second \euve catalog (Bowyer et al. 1996) was detected
independently in three \euve bands -- 100 \AA, 200 \AA, and 400 \AA.
Even though there are 89 stars within $3.5'$ of the position of this 
low-latitude ($b=+12^\circ$)
source, Fig. 1 shows that none is an obvious candidate. 
Even given the scatter of real stars around the schematic main-sequence line we have
plotted, the nearest of the stars, if they are main sequence dwarfs, are at a distance
of over 200 pc, making them unlikely EUV sources. This is especially true
given the 400 \AA\ detection, which implies a small distance to the source,
given the large ISM absorption cross-section in this band. If the stars
in the field are
white-dwarfs, they are too red (and hence
too cool) to be EUV sources.
The EUV source must therefore be an X-ray-quiet CV, or some new type of EUV source
that is optically faint or variable, or that has the colour and magnitude
of a distant late-type star but is closer than $\sim 100$ pc.

\subsubsection{EUVE 0807$+$210}
This source was listed as a 200 \AA\  ``Deep Survey'' detection in the 
first \euve source catalog (Bowyer et al. 1995), but is not listed
in the second catalog (Bowyer et al. 1996).
Among the 23 stars within $3.5'$ of the \euve position,
there is one (No. 23 in Table 3) that could be a nearby active late-type dwarf. As
seen in Fig. 2, it is an 11th mag star with $B-R=2$, consistent with
a K dwarf at a distance of $\sim 100$ pc. Given the scatter of real stars around
the main sequence, it could also be as close 50 pc. 
Its position is $3.2'$ from the \euve position, while the 90\% positional
error radius is typically $2.1'$. In a search of the \rosat
X-ray All-Sky-Survey Bright Source Catalog (Voges et al. 1996),
 we have found a 0.1--2.4 keV X-ray
source $0.65'$ north-east of this star (and hence consistent,
within the $0.5'$ positional accuracy). Optical spectroscopy can reveal whether
this star has characteristics of an active late-type star,
in which case it is likely to be the EUV source.
   If it turns out {\it not}
to be the EUV source then
there are no other good candidates in the field, and this is another case
of an EUV source with no clear counterparts among the two main
 classes of EUV emitters.

\subsubsection{RE 0847$+$594}
The \rosat 100 \AA\ detection of this source is at the $4.5\sigma$
level, below the $5.5\sigma$ threshold for inclusion in the second
\rosat WFC catalog (Pye et al. 1995). The \rosat data contain a positive
signal having a signal consistent with that expected from a point source,
but the expected number of spurious $4.5\sigma$ detections over the whole
sky is of order 200 (Pye et al. 1995). This source may therefore
be a false EUV detection.
There are 14 sources within $3.5'$ of the \rosat position, none of
which appear  to be a suitable EUV source candidate (Fig. 3).
We note that there is a radio source $82''$ north of the \rosat position,
with a 6 cm flux of 41 mJy and spectral slope $-0.9$ (Gregory \& Condon 1991), possibly 
associated with objects No. 2 or 5 in Table 4.

\subsubsection{RE 0922+710}
An EUV source was detected in this direction by both \rosat (at 150 \AA;
Pounds et al. 1993)
and \euve (at 400 \AA; Bowyer et al. 1994). It is not included in
the second \euve catalog (Bowyer et al. 1996). The \rosat
detection is at the $3.9\sigma$ level, below the $5.5\sigma$ threshold for
 inclusion in the second \rosat
catalog (Pye et al. 1995). 
There are 18 objects within $3.5'$ of the \rosat position.
Based on their positions on the colour-magnitude diagram (Fig. 4), none
is a suitable EUV source. Maoz et al. (1996) noted that
there is a 6 cm radio source and a faint \rosat PSPC X-ray source coincident
with the bluest object in the field, having $B-R=0.4$ mag, $R=18.3$ mag,
and lying within the error circles of both the \euve and \rosat
detections (object No. 3 in Table 5).
 They used optical spectroscopy to identify this object as a
redshift $z=2.432$ quasar. A 21 cm H I emission measurement in this
direction with a $21'$ beam showed that the total H I column density
was several orders of magnitude too high for the quasar to be 
the EUV source (see Maoz et al. 1996, for details).
The EUV source in this field therefore remains unidentified, with
no candidates among the main classes of EUV sources.
A recent repeated observation of this source with \euve (to be 
described elsewhere) confirms its reality.

\subsubsection{EUVE 1636$-$285}
This is a strong 600 \AA\ source appearing in both the first and second
\euve catalogs (Bowyer et al. 1994;1996).
Among the 95 sources within $3.5'$ of the \euve position,
the most plausible candidate
appears to be star No. 52 in Table 6, an $R=13.3$ mag, $B-R=2.9$ mag object
$2.6'$ from the \euve position. It could be an M star at $\sim100$ pc. (See Fig. 5).
Four other objects in Fig. 5 with $B-R\approx 3$ could also
be M stars at $\sim 100$ pc and so could conceivably be the EUV source.
On the other hand, one would expect a 600 \AA\ source to be
much nearer than 100 pc in order to escape ISM absorption.
Furthermore, one would expect an active late-type star
that is so strong at 600 \AA\ to have
been detected in the shorter-wavelength \euve bands as well.
We therefore suspect that none of the detected stars are 
late-type star or white-dwarf EUV sources.
 We fail to detect
this object in a repeated \euve observation  at 100 \AA\ (to be 
described elsewhere).

\subsubsection{EUVE 2053$-$175}
This source was listed as an unidentified Deep Survey 100~\AA\ emitter in the 
first \euve catalog (Bowyer et al. 1994), and as an identified
$V=10.4$ mag K star in the 
second catalog (Bowyer et al. 1996). The identification is described
by Craig et al. (1996b), who note it is not conclusive, since no
evidence of chromospheric activity or EUV variability are seen.
Prior to learning of this
identification in the second catalog, we analysed
the 38 stars in the short optical exposures within $3.5'$ of the \euve position.
We noted that the same star studied by Craig et al. (1996b) (their star ``No. 3'',
star No. 5 in Table 7 in this paper), $82''$ from the \euve position
 in the first \euve catalog,
 is a likely candidate EUV source. It has $R=9.7$ mag
and $B-R=1.3$ mag, consistent with a late G star at 100 pc (see Fig. 6).
 In a search of the \rosat
X-ray All-Sky-Survey Bright Source Catalog, we have found a 0.1--2.4 keV X-ray
source $15''$ north-west of this star (and hence consistent,
within the $0.5'$ positional accuracy). This strengthens the case that it is the EUV source.
Star No. 6 in Table 7, separated $91''$ from the \euve position,
may be a 13-mag K star at $\sim 100$ pc, and so is also
a plausible 
candidate. Craig et al. (1996b) obtained spectra for two additional
stars, No. 3 and No. 4 in Table 7, and assign them spectral types dK1
and dK7, respectively. Our method shows that neither are plausible
EUV source candidates, since they are too distant.

\subsubsection{EUVE 2114+503}
This is a strong source in the 1st \euve catalog (Bowyer et al. 1994).
It is omitted from the 2nd \euve catalog (Bowyer et al. 1996) due to improper
subtraction of background from an adjacent bright white dwarf
(A. Fruscione, personal communication). The 100 \AA\ detection is,
however, highly significant, and a recent determination of its
 parameters is given in
Craig et al. (1996b).
Our deep (3000 s in each band) exposures of this
crowded Galactic-plane field ($b=+1^\circ$) show 169 optical sources within
$3.5'$ of the \euve position.
 As seen in Fig. 7, about 20 of these
stars have colours and magnitudes consistent with nearby M stars that
could be the EUV source. The stars in this direction are almost
certainly shifted to this position in the colour-magnitude diagram
 by strong reddening, so it is also possible that {\it none} of them
is the EUV source. Indeed, Craig et al. (1996b) have obtained spectra
of nine stars in the field brighter than $V=14$ mag, none of
which are suitable EUV candidates. Either way, we cannot rule out,  based on the 
current data, that the EUV source in this field is a late-type star.
Note also the unusually (relative to the others) blue star (No. 78 in Table 8)
 with $B-R=0.14$ mag,
$R=19$ mag, $132''$ from the \euve position. While its place
 on the diagram argues against its
being a white-dwarf EUV source (it is too red and too distant),
 its colour suggests it is much closer than the other stars in the field,
and it deserves further study. It has not been observed by Craig et al. (1996b).
 A recent repeated observation
of this field with \euve (to be described elsewhere) does not 
reveal any EUV sources, suggesting the source may be a transient
EUV emitter.

\section{Summary}
We have presented $B$ and $R$ measurements of stars
in the fields of seven unidentified EUV sources detected
by \rosat and \euve. Using colour-magnitude diagrams,
we have argued that, in most of these fields, there
are no main-sequence stars near enough to
be the EUV source, nor white dwarfs that are hot  and near enough.
{\it Many of the unidentified EUV sources are therefore
not simply fainter members of these two classes of objects,
which are responsible for 90\% of identified EUV detections.}
 The unidentified EUV sources we have studied cannot be
X-ray binaries or AGNs, due to the absence of X-ray detections
in the \rosat X-ray All Sky Survey. It is possible that some
or all of the sources we have studied are false detections,
although several of them are quite strong, or have been detected
in several bands.

We conclude that, if they are real, the objects behind the EUV emission
in these fields must be either X-ray quiet CVs or a
new class of objects. If they are CVs, they are unusual, since
most CVs are X-ray sources. If they constitute a new class, then
 these objects are either very faint optically,
or are an unknown type of nearby Galactic star that mimics the colour
and magnitude of a distant late-type dwarf. In the former case,
the EUV/optical flux ratio,
 $\nu F_\nu(150{\rm \AA})/\nu F_\nu (6500{\rm\AA})\gtorder 100$, i.e. the
sources are some kind of extremely hot object, e.g. an isolated
old neutron star (see \S 1). The luminosities of the EUV sources 
discussed here are $\sim 10^{28}-10^{31}$ erg s$^{-1}$ (assuming
a distance of 10--100 pc), also similar
to those predicted for old neutron stars (e.g Madau \& Blaes 1994; Shemi 1995a).
 A third 
alternative is that the sources emit transiently, and were
``turned off'' at the time of the optical observations.
Obviously, these unidentified EUV sources warrant further study.

{\bf Acknowledgements}
We thank K. Anderson, N. Craig, A. Fruscione, P. Madau, 
T. Mazeh, A. Sternberg, and X. Wu for
helpful discussions and suggestions, M. Gardosh
for assistance with the observations, and the
referee, J. Pye, for constructive and useful comments.
 Astronomy at Wise Observatory
 is supported by grants from the Israel Academy of Science.
This work was supported by NASA-\euve grant NAG5-2913,
and has made use of the NASA/IPAC 
Extragalactic Database (NED), which is operated by JPL, Caltech, under
contract with NASA, the SIMBAD database which is operated at CDS Strasbourg,
the HEASARC database at NASA--GSFC, and the \rosat archive at MPE-Garching.

\begin{figure} 
   \centering \epsfxsize=320pt
   \epsfbox{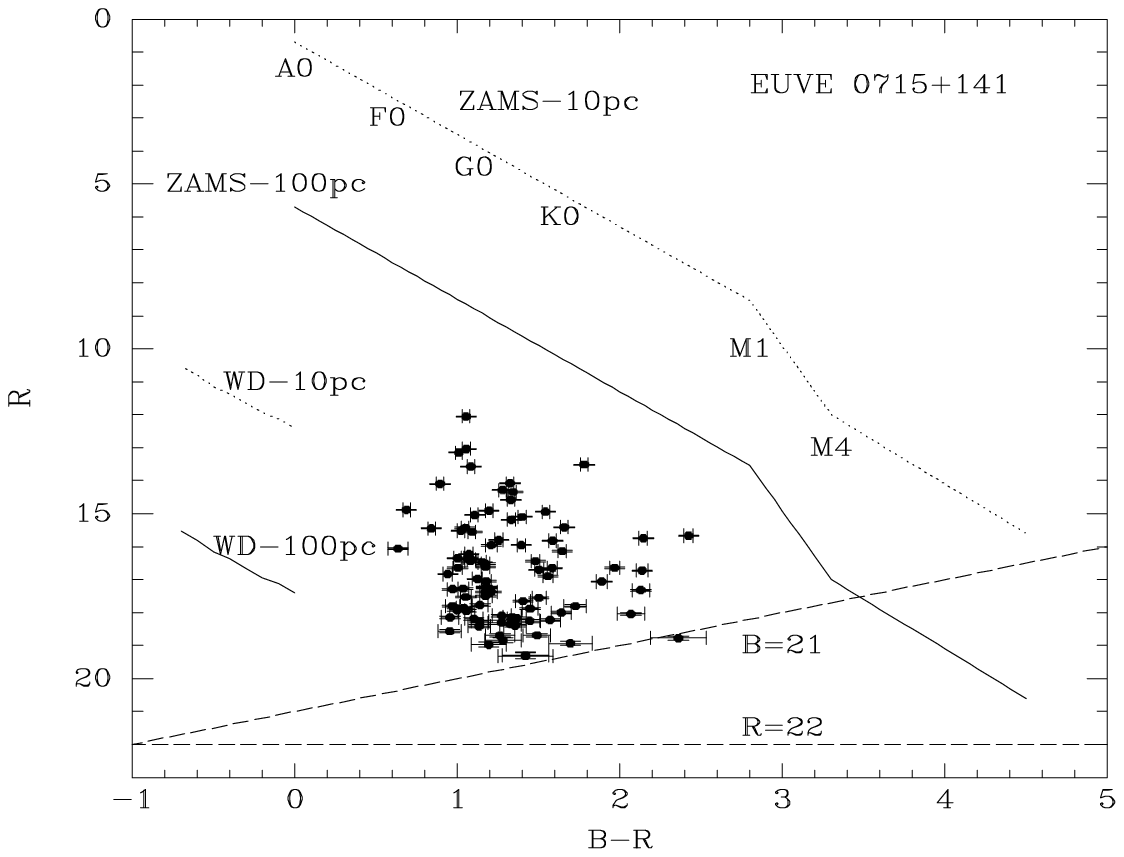}
   \caption{(and Figs. 2--7)
Colour-magnitude diagrams of $R$ vs. $B-R$ magnitude
 for stars within $3.5'$ of each of the EUV source
positions. The curves marked ZAMS -- 10 pc and ZAMS -- 100 pc show the
approximate locations of the zero-age main sequence at these distances,
with some of the spectral types indicated.
The curves marked WD -- 10 pc and WD -- 100 pc show the white dwarf
sequence at these distances, for white dwarfs bluer than $B-R=0$.
The dashed lines mark the detection limits, in $B$ and $R$, of the
optical exposures. To be EUV source candidates, white dwarfs need
to be bluer than $B-R=0$ (see Fig. 8), and main sequence dwarfs need to be
nearer than $\sim 100$ pc, i.e., above the ZAMS -- 100 pc curve.
See text for details.}
\label{f:0715} 
\end{figure} 

\begin{figure} 
   \centering \epsfxsize=320pt
   \epsfbox{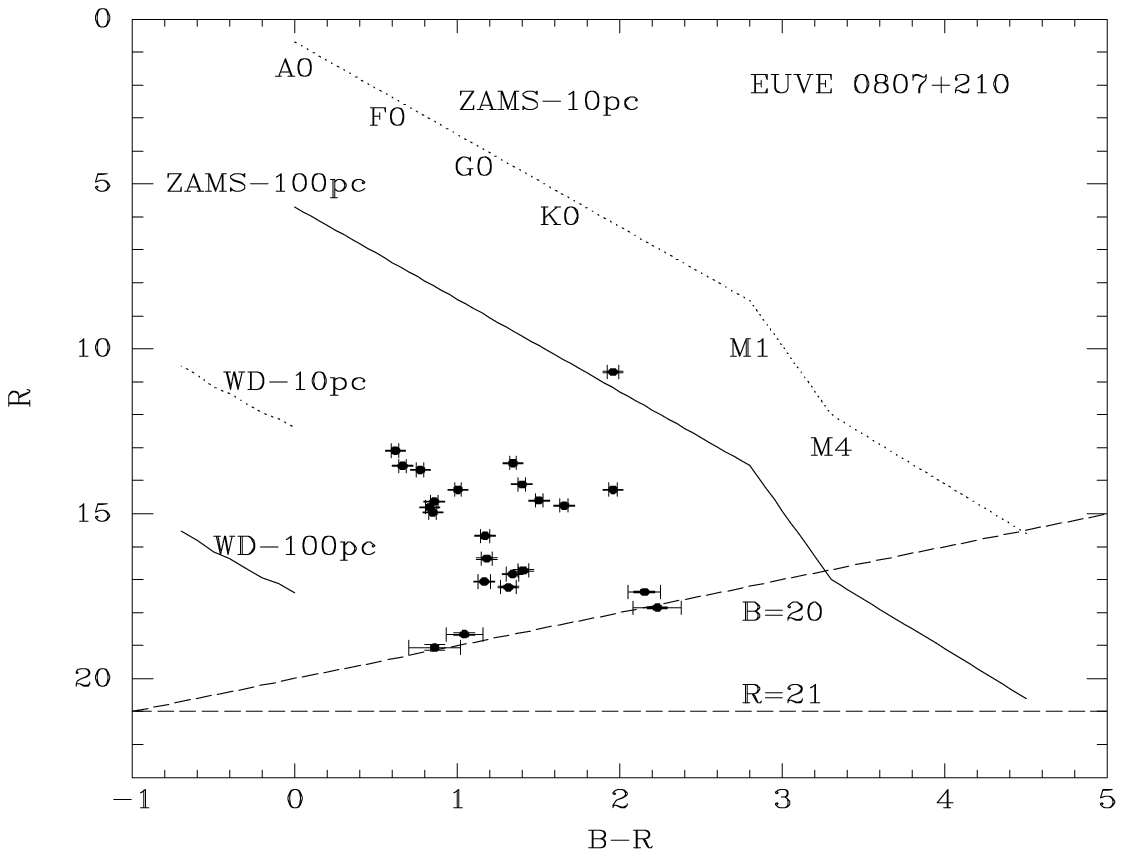}
\caption{}
\label{f:0807} 
\end{figure} 

\begin{figure} 
   \centering \epsfxsize=320pt
   \epsfbox{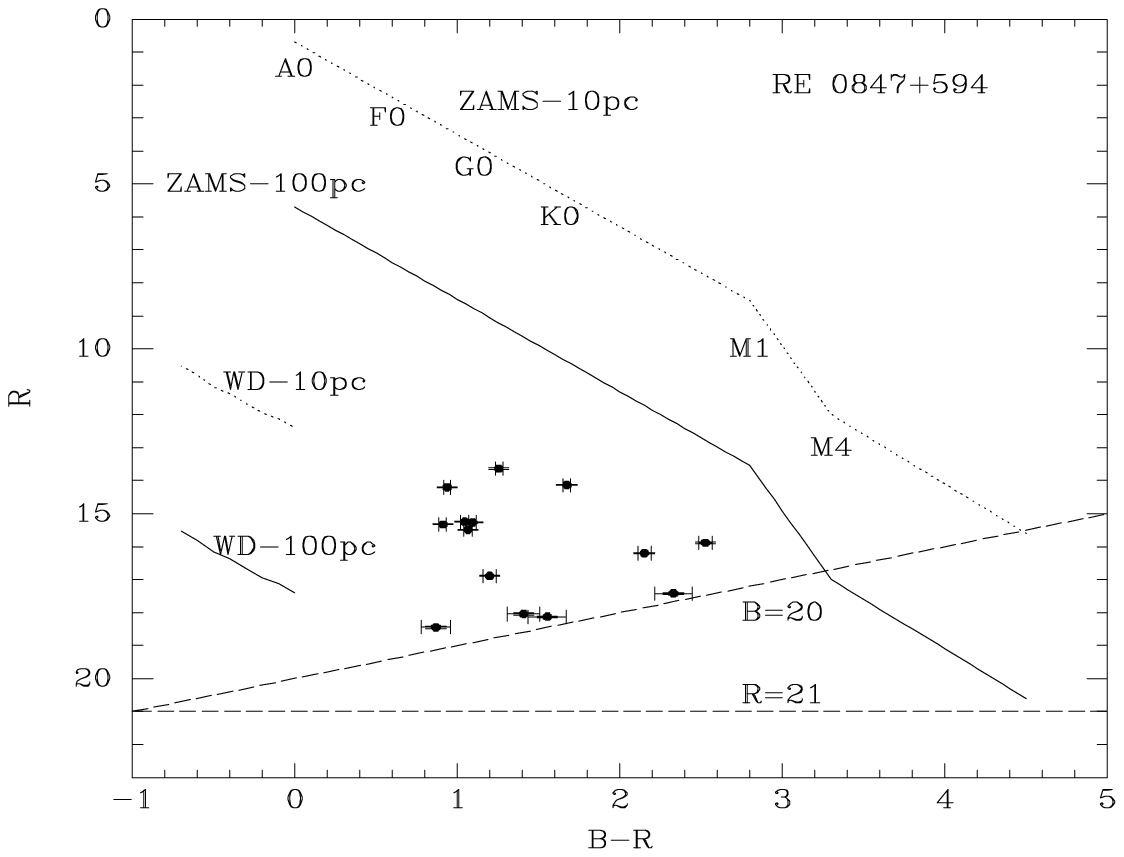}
\caption{}
\label{f:0847} 
\end{figure} 

\begin{figure} 
   \centering \epsfxsize=320pt
   \epsfbox{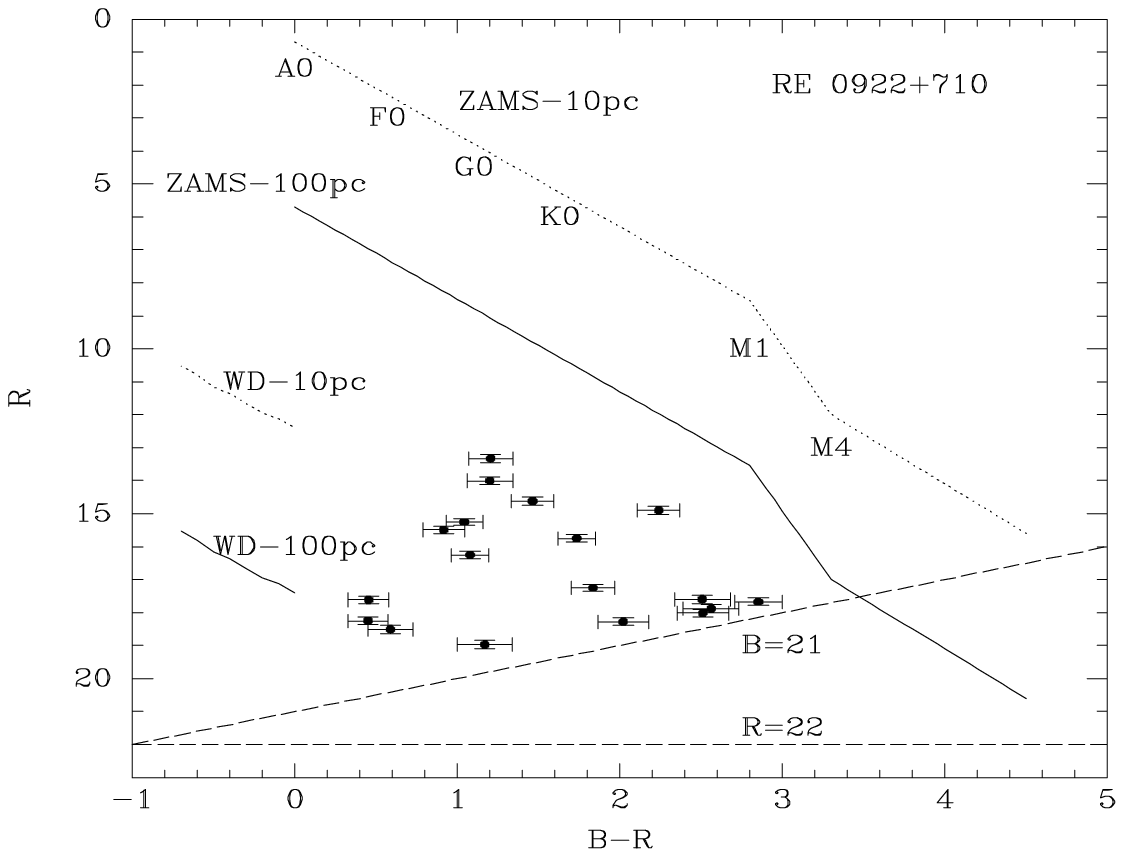}
\caption{}
\label{f:0922} 
\end{figure} 

\begin{figure} 
   \centering \epsfxsize=320pt
   \epsfbox{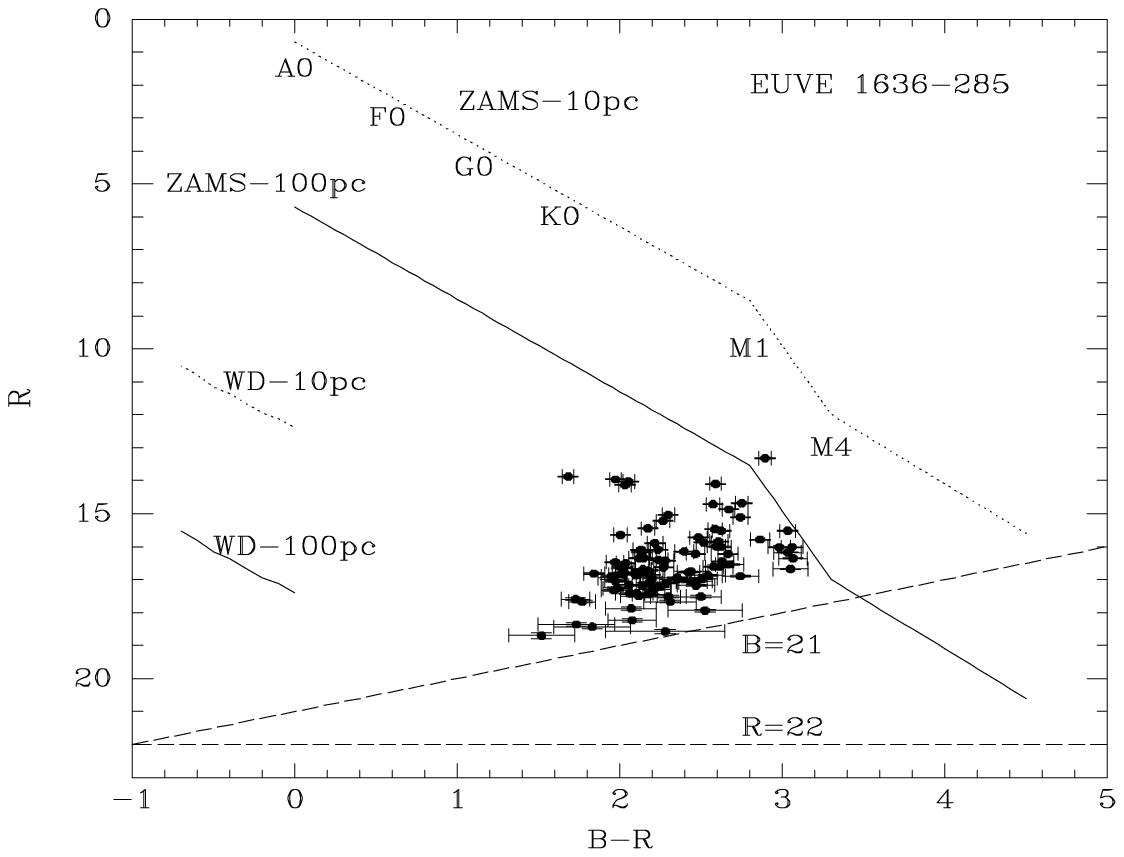}
\caption{}
\label{f:1636} 
\end{figure} 

\begin{figure} 
   \centering \epsfxsize=320pt
   \epsfbox{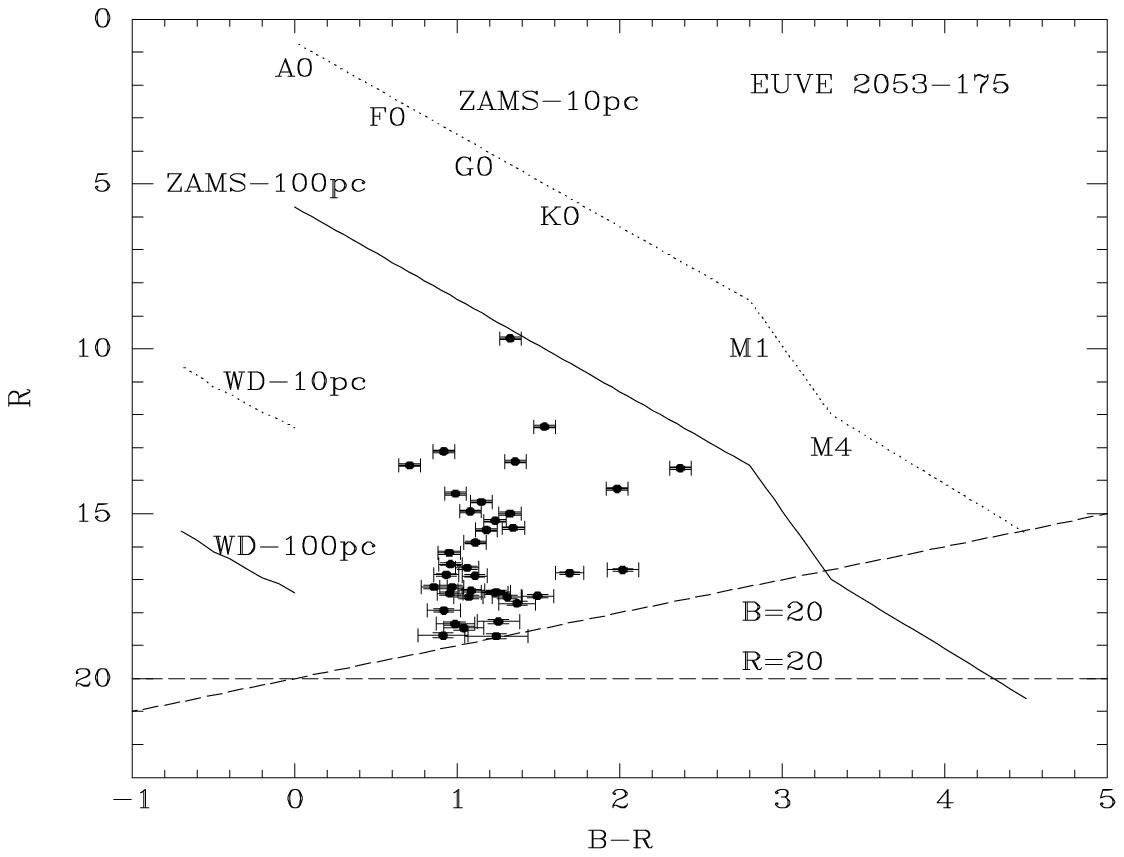}
\caption{}
\label{f:2053} 
\end{figure} 

\begin{figure} 
   \centering \epsfxsize=320pt
   \epsfbox{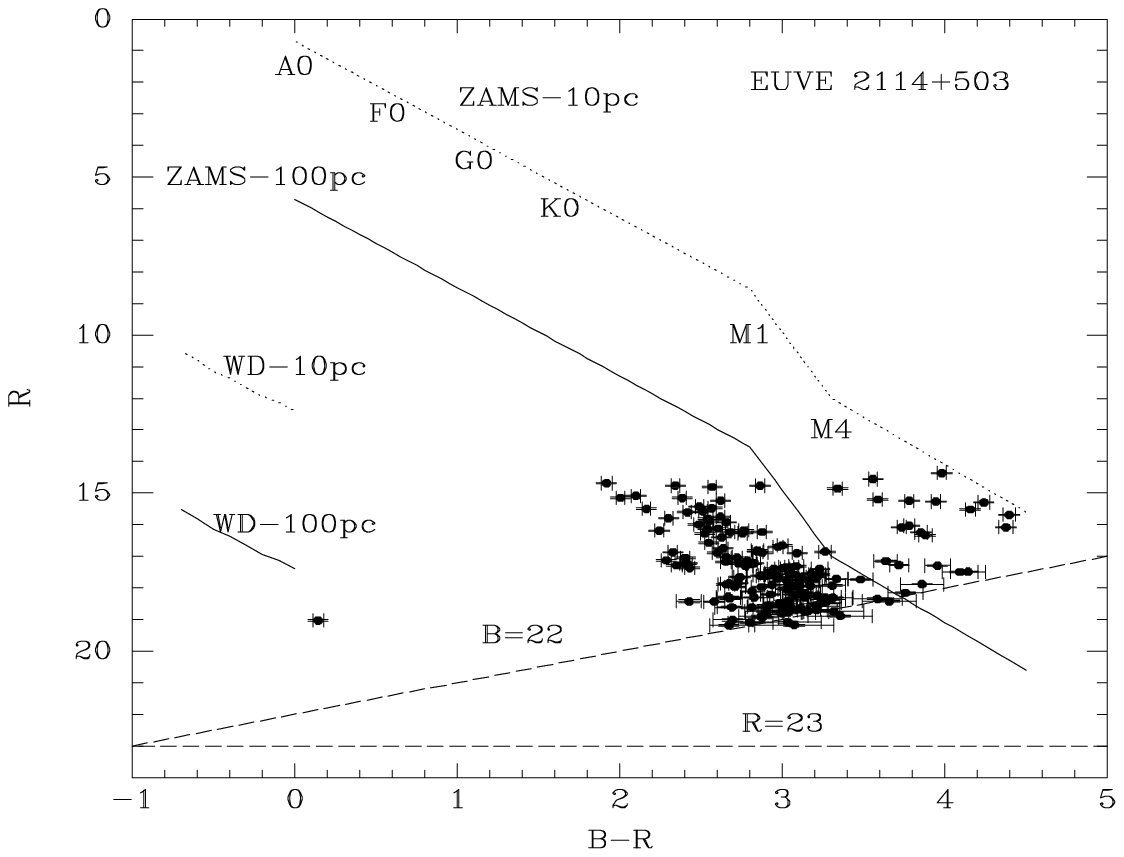}
\caption{}
\label{f:2114} 
\end{figure}

\begin{figure} 
   \centering \epsfxsize=320pt
   \epsfbox{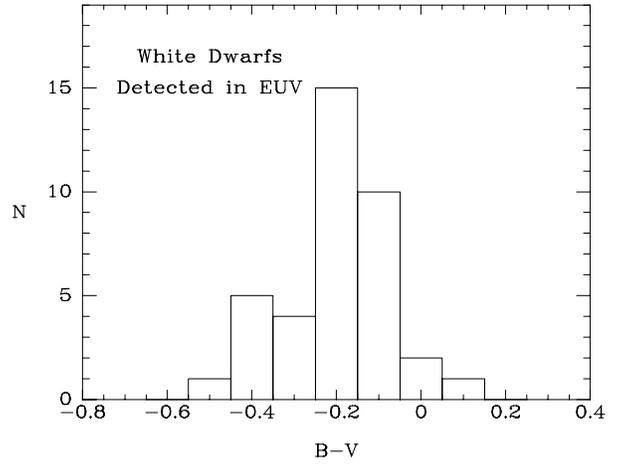}
   \caption{
Distribution of $B-V$ colour for all white dwarfs with this parameter
known, that have been detected in the EUV by \rosat or \euve.
Only hot white dwarfs, with $B-V\le 0.13$, have been detected in the EUV.}
\label{f:wd} 
\end{figure}

\end{document}